\documentclass[twocolumn,eqsecnum,prb]{revtex4}
\usepackage{amsfonts}
\usepackage[T1]{fontenc}
\usepackage{amsmath,amsbsy,amssymb,graphicx}

\let\mathbf=\boldsymbol

\begin{document}

\title{{\Large Quasi-Ferromagnet Spintronics\ in Graphene Nanodisk-Lead System}}
\author{Motohiko Ezawa}
\affiliation{Department of Applied Physics, University of Tokyo, Hongo 7-3-1, 113-8656,
Japan }

\begin{abstract}
A zigzag graphene nanodisk can be interpreted as a quantum dot with an
internal degree of freedom. It is well described by the infinite-range
Heisenberg model. We have investigated its thermodynamical properties. There
exists a quasi-phase transition between the quasi-ferromagnet and
quasi-paramagnet states, as signaled by a sharp peak in the specific heat
and in the susceptability. We have also analyzed how thermodynamical
properties are affected when two leads are attached to the nanodisk. It is
shown that lead effects are described by the many-spin Kondo Hamiltonian.
There appears a new peak in the specific heat, and the multiplicity of the
ground state becomes just one half of the system without leads. Another lead
effect is to enhance the ferromagnetic order. Being a ferromagnet, a
nanodisk can be used as a spin filter. Furthermore, since the relaxation
time is finite, it is possible to control the spin of the nanodisk by an
external spin current. We then propose a rich variety of spintronic devices
made of nanodisks and leads, such as spin memory, spin amplifier, spin
valve, spin-field-effect transistor, spin diode and spin logic gates such as
spin-XNOR gate and spin-XOR gate. Graphene nanodisks could well be basic
components of future nanoelectronic and spintronic devices.
\end{abstract}

\maketitle


\address{{\normalsize Department of Applied Physics, University of Tokyo, Hongo
7-3-1, 113-8656, Japan }}

\section{Introduction}

Graphene nanostructure\cite{GraphExA,GraphExB,GraphExC} has attracted much
attention for its potential for future application in nanoelectronics and
spintronics. In particular, in graphene nanoribbons\cite%
{Fujita,EzawaPRB,Brey,Rojas,Son,Barone,Kim} the low-energy bands are almost
flat at the Fermi level due to the edge states. Such a peculiar band
structure has motivated many researchers to investigate their electronic and
magnetic properties.

Another basic element of graphene derivatives is a graphene nanodisk\cite%
{EzawaDisk,EzawaPhysica,Fernandez,Hod,Wang,EzawaCoulomb,Wang2}. It is a
nanometer-scale disk-like material which has a closed edge. There are many
type of nanodisks, where typical examples are displayed in Fig.\ref%
{FigNanodisk}. Among them, the trigonal zigzag nanodisk is prominent in its
electronic and magnetic properties because there exist $N$-fold degenerated
half-filled zero-energy states when its size is $N$. Furthermore, spins make
a strong ferromagnetic order due to the exchange interaction as large as the
Coulomb interaction\cite{EzawaDisk}, and hence the relaxation time is finite
but quite large even if the size $N$ is very small. We have called such a
system quasi-ferromagnet\cite{EzawaDisk}.

\begin{figure}[t]
\centerline{\includegraphics[width=0.4\textwidth]{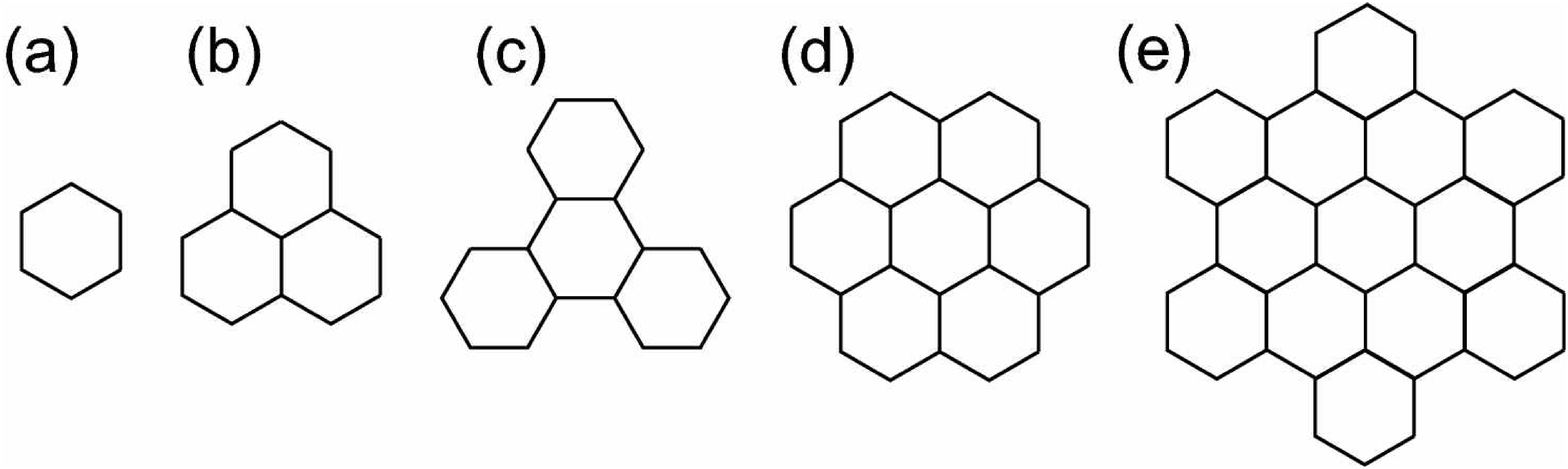}}
\caption{Basic configurations of typical graphene nanodisks. (a) Benzene.
(b) Trigonal zigzag nanodisk (phenalene). (c) Trigonal armchair nanodisk
(triphenylene). (d) Hexagonal zigzag nanodisk (coronene). (e) Hexagonal
armchair nanodisk (hexa benzocoronene)\protect\cite{Rader}. }
\label{FigNanodisk}
\end{figure}

In this paper we make a further study of quasi-ferromagnet by exploring
thermodynamical properties of the nanodisk-spin system. The system is well
described by the infinite-range Heisenberg model. A nanodisk is interpreted
as a quantum dot with an internal degree of freedom. It is exactly solvable.
Constructing the partition function, we calculate the specific heat, the
entropy, the magnetization and the susceptibility. We find a sharp peak in
the specific heat and in the susceptibility, which is interpreted as a
quasi-phase transition between the quasi-ferromagnet and quasi-paramagnet
states.

We then investigate a nanodisk-lead system, where the lead is made of a
zigzag graphene nanoribbon or an ordinary metallic wire. (We refer to it as
a graphene lead or a metallic lead for brevity.) We perform the
Schrieffer-Wolff transformation to derive the many-spin Kondo Hamiltonian
describing the lead effects. Constructing the partition function, we analyze
thermodynamical properties. There appears a new peak in the specific heat
but not in the susceptibility for small size nanodisks, $N\lesssim 10$. Near
the peak the internal energy is found to decrease. We show in the instance
of the metallic lead that the band width of free electrons in the lead
becomes narrower due to the Kondo coupling. We interpret this phenomenon to
mean that some free electrons in the lead are consumed to make spin coupling
with electrons in the nanodisk. Furthermore, the multiplicity of the ground
state becomes just one half of that in the system without leads. They are
indications of Kondo effects due to the Kondo interaction between electrons
in the lead and the nanodisk.

With respect to the ferromagnetic order, we find that the lead effect is to
enhance the order. This is an important property to fabricate spintronic
circuits by connecting leads to nanodisks in nanodevices. Being a
ferromagnet, a nanodisk can be used as a spin filter. Furthermore, since the
relaxation time is finite, it is possible to control the spin of the
nanodisk by an external spin current. We propose a rich variety of
spintronic devices made of nanodisks and leads, such as spin memory, spin
amplifier, spin valve, spin-field-effect transistor, spin diode and spin
logic gates such as spin-XNOR gate and spin-XOR gate. Graphene nanodisks
could well be basic components of future nanoelectronic and spintronic
devices.

This paper is organized as follows. In Sec.\ref{SecNanodisk}, we summarize
the basic notion of trigonal zigzag nanodisks. The low-energy physics is
described by electrons in the $N$-fold degenerated zero-energy sector, which
form a quasi-ferromagnet due to large exchange interactions. We also analyze
thermodynamical properties of the nanodisk-spin system. In Sec.\ref{SecKondo}%
, we derive the many-spin Kondo Hamiltonian by the Schrieffer-Wolff
transformation in the nanodisk-spin system coupled with graphene leads and
also metallic leads. The partition function is calculated in the two steps.
First, we perform a functional integration over the lead-electron degree of
freedom. Second, we sum up over the nanodisk-spin degree of freedom. We then
analyze thermodynamical properties of the nanodisk-lead system. In Sec.\ref%
{SecApplication}, we propose some spintronics devices made of nanodisks and
leads.

\section{Nanodisk Quasi-Ferromagnets}

\label{SecNanodisk}

\subsection{Zero-Energy Sector}

\begin{figure}[t]
\centerline{\includegraphics[width=0.5\textwidth]{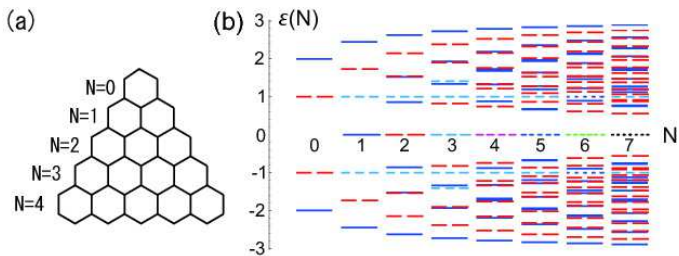}}
\caption{(a) Geometric configuration of trigonal zigzag nanodisks. It is
convenient to introduce the size parameter $N$ in this way. The $0$-trigonal
nanodisk consists of a single Benzene, and so on. The number of carbon atoms
are related as $N_{\text{C}}=N^{2}+6N+6$. (b) Density of states of the $N$%
-trigonal nanodisk for $N=0,1,2,\cdots ,7$. The horizontal axis is the size $%
N$ and the vertical axis is the energy $\protect\varepsilon (N)$ in units of 
$t=2.7$eV. Dots on colored bar indicate the degeneracy of energy levels.}
\label{FigDotSamp}
\end{figure}

We calculate the energy spectra of graphene derivatives based on the
nearest-neighbor tight-binding model, which has been successfully applied to
the studies of carbon nanotubes and nanoribbons. The Hamiltonian is defined
by%
\begin{equation}
H=\sum_{i}\varepsilon _{i}c_{i}^{\dagger }c_{i}+\sum_{\left\langle
i,j\right\rangle }t_{ij}c_{i}^{\dagger }c_{j},  \label{HamilTB}
\end{equation}%
where $\varepsilon _{i}$ is the site energy, $t_{ij}$ is the transfer
energy, and $c_{i}^{\dagger }$ is the creation operator of the $\pi $
electron at the site $i$. The summation is taken over all nearest
neighboring sites $\left\langle i,j\right\rangle $. Owing to their
homogeneous geometrical configuration, we may take constant values for these
energies, $\varepsilon _{i}=\varepsilon _{\text{F}}$ and $t_{ij}=t\approx
2.70$eV. Then, the diagonal term in (\ref{HamilTB}) yields just a constant, $%
\varepsilon _{\text{F}}N_{\text{C}}$, where $N_{\text{C}}$ is the number of
carbon atoms in the system. The Hamiltonian (\ref{HamilTB}) yields the Dirac
electrons for graphene\cite{GraphExA,GraphExB,GraphExC}. There exists one
electron per one carbon and the band-filling factor is 1/2. It is customary
to choose the zero-energy level of the tight-binding Hamiltonian (\ref%
{HamilTB}) at this point so that the energy spectrum is symmetric between
the positive and negative energy states. Therefore, the system is metallic
provided that there exists zero-energy states in the spectrum. A comment is
in order. It is understood that carbon atoms at edges are terminated by
hydrogen atoms. We carry out the calculation\cite{EzawaPRB,EzawaDisk}
together with this condition.

It is straightforward to derive the energy spectrum $E_{i}$ together with
its degeneracy $g_{i}$ for each nanodisk by diagonalizing the Hamiltonian (%
\ref{HamilTB}). The density of state is given by%
\begin{equation}
D\left( \varepsilon \right) =\sum_{i=1}^{N_{\text{C}}}g_{i}\delta \left(
\varepsilon -E_{i}\right) .  \label{DOS}
\end{equation}%
We have found that the emergence of zero-energy states is surprisingly rare.
Among typical nanodisks, only trigonal zigzag nanodisks have degenerate
zero-energy states and show metallic ferromagnetism. As an example we
display the density of state (\ref{DOS}) of trigonal zigzag nanodisks in Fig.%
\ref{FigDotSamp}. We have classified them by the size parameter $N$ as
defined in Fig.\ref{FigDotSamp}(a), in terms of which the number of carbons
is given by $N_{\text{C}}=N^{2}+6N+6$.

\begin{figure}[t]
\centerline{\includegraphics[width=0.18\textwidth]{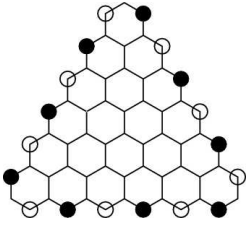}}
\caption{The zero-energy states of the trigonal nanodisk with size $N=5$.
There are $5$ degenerate states. The solid (open) circles denote the
amplitude $\protect\omega _{i}$ are positive (negative). The amplitude is
proportional to the radius of circle. Electrons are entirely localized on
edges in one of the states. }
\label{FigWave}
\end{figure}

The size-$N$ nanodisk has $N$-fold degenerated zero-energy states, where the
gap energy is as large as a few eV. Hence it is a good approximation to
investigate the electron-electron interaction physics only in the
zero-energy sector, by projecting the system to the subspace made of those
zero-energy states. The zero-energy sector consists of $N$ orthonormal
states $|f_{\alpha }\rangle $, $\alpha =1,2,\cdots ,N$, together with the SU(%
$N$) symmetry. We can expand the wave function of the state $|f_{\alpha
}\rangle $ as%
\begin{equation}
f_{\alpha }(\boldsymbol{x})=\sum_{i}\omega _{i}^{\alpha }\varphi _{i}(%
\boldsymbol{x}),  \label{EqA}
\end{equation}%
where $\varphi _{i}(\boldsymbol{x})$ is the Wannier function localized at
the site $i$. The amplitude $\omega _{i}^{\alpha }$ is calculable by
diagonalizing the Hamiltonian (\ref{HamilTB}). All of them are found to be
real. It is intriguing that one of the wave functions is entirely localized
on edge sites for the nanodisk with $N=$odd, as in Fig.\ref{FigWave}, where
the solid (open) circles denote the amplitude $\omega _{i}$ are positive
(negative). The amplitude is proportional to the radius of circle. Such a
wave function does not exist for the nanodisk with $N=$even.

\subsection{Coulomb Interactions}

We include the Coulomb interaction between electrons in the zero-energy
sector\cite{EzawaCoulomb}. It is straightforward to rewrite the Coulomb
Hamiltonian as $H_{\text{D}}=H_{\text{S}}+H_{\text{U}}$ with 
\begin{subequations}
\begin{align}
H_{\text{S}}=& -2\sum_{\alpha >\beta }J_{\alpha \beta }\mathbf{S}(\alpha
)\cdot \mathbf{S}(\beta ),  \label{HamilFerro} \\
H_{\text{U}}=& \sum_{\alpha >\beta }\left( U_{\alpha \beta }-\frac{1}{2}%
J_{\alpha \beta }\right) n\left( \alpha \right) n\left( \beta \right)
+\sum_{\alpha }U_{\alpha \alpha }n\left( \alpha \right) ,
\end{align}%
\end{subequations}
where $U_{\alpha \beta }$ and $J_{\alpha \beta }$ are the Coulomb energy and
the exchange energy between electrons in the states $|f_{\alpha }\rangle $
and $|f_{\beta }\rangle $. Here, $n\left( \alpha \right) $ is the number
operator and $\mathbf{S}(\alpha )$ is the spin operator, 
\begin{subequations}
\begin{align}
n\left( \alpha \right) & =\sum_{\sigma }d_{\sigma }^{\dag }(\alpha
)d_{\sigma }(\alpha ), \\
\mathbf{S}(\alpha )& =\frac{1}{2}\sum_{\sigma \sigma ^{\prime }}d_{\sigma
}^{\dag }(\alpha )\mathbf{\tau }_{\sigma \sigma ^{\prime }}d_{\sigma
^{\prime }}(\alpha ),
\end{align}%
\end{subequations}
with $d_{\sigma }(\alpha )$ the annihilation operator of electron with spin $%
\sigma =\uparrow ,\downarrow $ in the state $|f_{\alpha }\rangle $; $\mathbf{%
\tau }$ is the Pauli matrix.

The remarkable feature is that there exists a large overlap between the wave
functions $f_{\alpha }(\boldsymbol{x})$ and $f_{\beta }(\boldsymbol{x})$, $%
\alpha \neq \beta $, since the state $|f_{\alpha }\rangle $ is an ensemble
of sites as in (\ref{EqA}) and identical sites are included in $|f_{\alpha
}\rangle $ and $|f_{\beta }\rangle $. Consequently, the dominant
contributions come from the on-site Coulomb terms not only for the Coulomb
energy but also for the exchange energy. Indeed, it follows that $U_{\alpha
\beta }=J_{\alpha \beta }$ in the on-site approximation. We thus obtain 
\begin{equation}
U_{\alpha \beta }\simeq J_{\alpha \beta }\simeq {U}\sum_{i}(\omega
_{i}^{\alpha }\omega _{i}^{\beta })^{2},  \label{EqY}
\end{equation}%
where%
\begin{equation}
U\equiv \int \!d^{3}xd^{3}y\;\varphi _{i}^{\ast }(\boldsymbol{x})\varphi
_{i}(\boldsymbol{x})V(\mathbf{x}-\mathbf{y})\varphi _{i}^{\ast }(\boldsymbol{%
y})\varphi _{i}(\boldsymbol{y}),
\end{equation}%
with the Coulomb potential $V(\mathbf{x}-\mathbf{y})$. The Coulomb energy $U$
is of the order of $1$eV because the lattice spacing of the carbon atoms is $%
\sim 1$\AA\ in graphene. Coulomb blockade peaks appear at $\mu =\varepsilon
_{\alpha }$ and $\mu =\varepsilon _{\alpha }+U_{\alpha \beta }$, where new
channels open\cite{EzawaCoulomb}. We can determine experimentally the energy 
$\varepsilon _{\alpha }$ and $U_{\alpha \beta }$ by identifying Coulomb
blockade peaks.

\subsection{SU($N$) Approximation}

Since the exchange energy $J_{\alpha \beta }$ is as large as the Coulomb
energy $U_{\alpha \beta }$, the spin stiffness $J_{\alpha \beta }$ is quite
large. Furthermore, we have checked\cite{EzawaCoulomb} numerically that all $%
J_{\alpha \beta }$ are of the same order of magnitude for any pair of $%
\alpha $ and $\beta $, implying that the SU($N$) symmetry is broken but not
so strongly in the Hamiltonian (\ref{HamilFerro}). It is a good
approximation to start with the exact SU($N$) symmetry. Then, the
zero-energy sector is described by the Hamiltonian $H_{\text{D}}=H_{\text{S}%
}+H_{\text{U}}$, with 
\begin{subequations}
\label{HamilD}
\begin{align}
H_{\text{S}}=& -J\sum_{\alpha \neq \beta }\mathbf{S}(\alpha )\cdot \mathbf{S}%
(\beta ), \\
H_{\text{U}}=& \left( \frac{U}{2}-\frac{J}{4}\right) \sum_{\alpha \neq \beta
}n\left( \alpha \right) n\left( \beta \right) +U\sum_{\alpha }n\left( \alpha
\right) ,
\end{align}%
\end{subequations}
where $J\approx U$. The term $H_{\text{S}}$ is known as the infinite-range
Heisenberg model. We rewrite them as 
\begin{subequations}
\begin{align}
H_{\text{S}}=& -J\mathbf{S}_{\text{tot}}^{2}+\frac{3}{4}Jn_{\text{tot}},
\label{HamilDS} \\
H_{\text{U}}=& \left( \frac{U}{2}-\frac{J}{4}\right) (n_{\text{tot}}^{2}+1),
\label{HamilDU}
\end{align}%
\end{subequations}
where $\mathbf{S}_{\text{tot}}=\sum_{\alpha }\mathbf{S}\left( \alpha \right) 
$ is the total spin, and $n_{\text{tot}}=\sum_{\alpha }n\left( \alpha
\right) $ is the total electron number.

The ground states of nanodisks are half filled. We restrict the Hilbert
space to the half-filling sector, 
\begin{equation}
n\left( \alpha \right) =n_{\uparrow }\left( \alpha \right) +n_{\downarrow
}\left( \alpha \right) =1.
\end{equation}%
The Hamiltonian (\ref{HamilDS}) is reduced to the Heisenberg model,%
\begin{equation}
H_{\text{S}}=-J\mathbf{S}_{\text{tot}}\cdot \mathbf{S}_{\text{tot}},
\end{equation}%
where we have neglected an irrelevant constant term, $(3/4)JN$. This is
exactly diagonalizable, $H_{\text{S}}|\Psi \rangle =E_{s}|\Psi \rangle $,
with%
\begin{equation}
E_{s}=-Js(s+1),
\end{equation}%
where $s$ takes values from $N/2$ down to $1/2$ or $0$, depending on whether 
$N$ is odd or even,%
\begin{equation}
s=\frac{N}{2},\frac{N}{2}-1,\frac{N}{2}-2,\cdots ,\qquad s\geq 0.
\end{equation}%
The Hilbert space is diagonalized, 
\begin{equation}
\mathbb{H}=\otimes ^{N}\mathbb{H}_{1/2}=\oplus ^{g_{N}\left( s\right) }%
\mathbb{H}_{s},
\end{equation}%
where $\mathcal{H}_{s}$ denotes the $\left( 2s+1\right) $ dimensional
Hilbert space associate with an irreducible representation of SU(2). The
multiplicities $g_{N}\left( s\right) $ satisfies the recursion relation
coming from the spin synthesizing rule,%
\begin{equation}
g_{N}\left( s\right) =g_{N-1}\left( s-\frac{1}{2}\right) +g_{N-1}\left( s+%
\frac{1}{2}\right) .
\end{equation}%
We solve this as 
\begin{equation}
g_{N}\left( \frac{N}{2}-q\right) =_{N}\!\!C_{q}-_{N}\!\!C_{q-1.}
\end{equation}%
The total degeneracy of the energy level $E_{s}$ is $\left( 2s+1\right)
g_{N}(s)$.

At half filling, the eigenstate of the Hamiltonian $H_{\text{D}}$ is labeled
as $\left\vert \Psi \right\rangle =\left\vert n_{\text{tot}%
},s,m\right\rangle $, where $s$ is the total spin and $m$ is its $z$%
-component. We refer to the total spin $\mathbf{S}_{\text{tot}}$ of a
nanodisk as the nanodisk spin.

\subsection{Thermodynamical Properties}

\label{SecStat1}

\begin{figure}[t]
\includegraphics[width=0.5\textwidth]{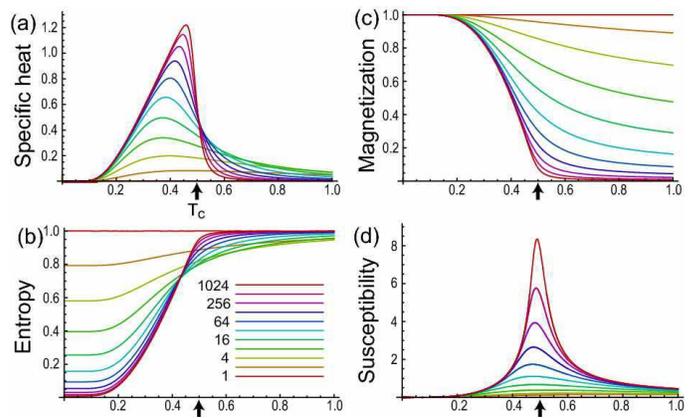}
\caption{Thermodynamical properties of the nanodisk-spin system. (a) The
specific heat $C$ in unit of $k_{\text{B}}N$. (b) The entropy $S$ in unit of 
$k_{\text{B}}N\log 2$. (c) The magnetization $\left\langle \mathbf{S}_{\text{%
tot}}^{2}\right\rangle $ in unit of $S_{g}^{2}$. (d) The susceptibility $%
\protect\chi $ in unit of $S_{g}$. The size is $N=1,2,2^{2},\cdots 2^{10}$.
The horizontal axis stands for the temperature $T$ in unit of $JN/k_{\text{B}%
}$. The arrow represents the phase transition point $T_{c}$ in the limit $%
N\rightarrow \infty $.}
\label{FigThermo}
\end{figure}

We have a complete set of the eigenenergies together with their
degeneracies. The partition function of the nanodisk with size $N$ is
exactly calculable,%
\begin{align}
Z_{\text{S}}& =\sum_{s}\left( 2s+1\right) g_{N}(s)e^{-\beta E_{s}}  \notag \\
& =\sum_{q=0}^{N/2}\left( N-2q+1\right) \left( _{N}C_{q}-_{N}C_{q-1}\right) 
\notag \\
& \times \exp \left[ \beta J\left( \frac{N}{2}-q\right) \left( \frac{N}{2}%
-q+1\right) \right] .
\end{align}%
According to the standard procedure we can evaluate the specific heat $C(T)$%
, the entropy $S(T)$, the magnetization $\left\langle \mathbf{S}_{\text{tot}%
}^{2}\right\rangle $ and the susceptibility $\chi =\frac{1}{k_{\text{B}}T}%
\left( \left\langle \mathbf{S}_{\text{tot}}^{2}\right\rangle -\left\langle 
\mathbf{S}_{\text{tot}}\right\rangle ^{2}\right) $ from this partition
function, where%
\begin{equation}
S_{g}=\sqrt{\frac{N}{2}\left( \frac{N}{2}+1\right) }
\end{equation}%
is the ground-state value of the total spin. The entropy is given by%
\begin{equation}
S\left( 0\right) =k_{\text{B}}\log (N+1)  \label{SN0}
\end{equation}%
at zero temperature. We display them in Fig.\ref{FigThermo} for size $%
N=1,2,2^{2},\cdots 2^{10}$.

There appear singularities in thermodynamical quantities as $N\rightarrow
\infty $, which represent a phase transition at $T_{c}$ between the
ferromagnet and paramagnet states,%
\begin{equation}
T_{c}=\frac{JN}{2k_{\text{B}}}.
\end{equation}
For finite $N$, there are steep changes around $T_{c}$, though they are not
singularities. It is not a phase transition. However, it would be reasonable
to call it a quasi-phase transition between the quasi-ferromagnet and
quasi-paramagnet states. Such a quasi-phase transition is manifest even in
finite systems with $N=100$ $\sim $ $1000$.

The specific heat and the magnetization take nonzero-values for $T>T_{c}$
[Fig.\ref{FigThermo}(a),(c)], which is zero in the limit $N\rightarrow
\infty $. The entropy for $T>T_{c}$ is lower than that of the paramagnet
[Fig.\ref{FigThermo}(b)]. These results indicate the existence of some
correlations in the quasi-paramagnet state. The maximum value of the
susceptibility increases linearly as $N$ becomes large. It is an indicator
of the quasi-phase transition.

\section{Many-Spin Kondo Effects}

\label{SecKondo}

We proceed to investigate how thermodynamical properties of the nanodisk is
affected by the attachment of the leads. The nanodisk is no longer in the
half-filled state, when charges transfer between the nanodisk and the leads.
However, the nanodisk remains to be half filled, when a charge transfers
from the lead to the nanodisk and then transfers back from the nanodisk to
the lead. Such a process is the second order effect in the tunneling
coupling constant $\tilde{t}$. We now show that it is described by the
many-spin Kondo Hamiltonian.

\begin{figure}[t]
\includegraphics[width=0.3\textwidth]{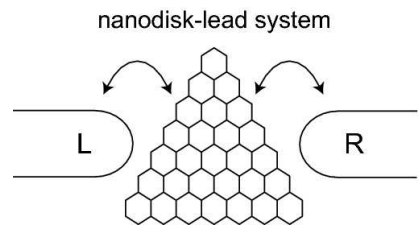}
\caption{The nanodisk-lead system. The nanodisk with $N=7$ is connected to
the right and left leads by tunneling coupling $t_{\text{R}}$ and $t_{\text{L%
}}$. }
\label{FigNLead}
\end{figure}

\subsection{Nanodisk-Lead System}

We analyze a system made of a nanodisk connected to two leads [Fig.\ref%
{FigNLead}]. The model Hamiltonian is given by%
\begin{equation}
H=H_{\text{S}}+H_{\text{U}}+H_{\text{L}}+H_{\text{TL}}+H_{\text{TR}}.
\label{TotalHamil}
\end{equation}%
Here, $H_{\text{L}}$ is the lead Hamiltonian, 
\begin{equation}
H_{\text{L}}=\sum_{k\sigma }\varepsilon \left( k\right) (c_{k\sigma }^{\text{%
L}\dagger }c_{k\sigma }^{\text{L}}+c_{k\sigma }^{\text{R}\dagger }c_{k\sigma
}^{\text{R}}),
\end{equation}%
describing a noninteracting electron gas in the leads with the dispersion
relation $\varepsilon \left( k\right) $, where $c_{k\sigma }^{\chi }$ is the
annihilation operator of electron with the wave number $k$ and the spin $%
\sigma $ in the left ($\chi =$L) or right ($\chi =$R) lead. On the other
hand, $H_{\text{TL}}$ and $H_{\text{TR}}$ are the transfer Hamiltonians
between the left (L) and right (R) leads and the nanodisk, respectively, 
\begin{subequations}
\label{HamilTLR}
\begin{align}
H_{\text{TL}}=& t_{\text{L}}\sum_{k\sigma }\sum_{\alpha }[c_{k\sigma }^{%
\text{L}\dagger }d_{\sigma }(\alpha )+d_{\sigma }^{\dagger }(\alpha
)c_{k\sigma }^{\text{L}}], \\
H_{\text{TR}}=& t_{\text{R}}\sum_{k\sigma }\sum_{\alpha }[c_{k\sigma }^{%
\text{R}\dagger }d_{\sigma }(\alpha )+d_{\sigma }^{\dagger }(\alpha
)c_{k\sigma }^{\text{R}}],
\end{align}%
\end{subequations}
with $t_{\chi }$ the tunneling coupling constant. We have assumed that the
spin does not flip in the tunneling process.

The nanodisk-lead system looks similar to that of the $N$-dot system.\cite%
{TaruchaRev} However, there exists a crucial difference. On one hand, in the
ordinary $N$-dot system, an electron hops from one dot to another dot. On
the other hand, in our nanodisk system, the index $\alpha $ of the
Hamiltonian runs over the $N$-fold degenerate states and not over the sites.
According to the Hamiltonian (\ref{HamilTLR}), an electron does not hop from
one state to another state. Hence, it is more appropriate to regard our
nanodisk as a one-dot system with an internal degree of freedom.

It is convenient to make the transformation 
\begin{equation}
\left( 
\begin{array}{c}
c_{k\sigma }^{\text{e}} \\ 
c_{k\sigma }^{\text{o}}%
\end{array}%
\right) =\frac{1}{\tilde{t}}\left( 
\begin{array}{cc}
t_{\text{L}}^{\ast } & t_{\text{R}}^{\ast } \\ 
-t_{\text{R}} & t_{\text{L}}%
\end{array}%
\right) \left( 
\begin{array}{c}
c_{k\sigma }^{\text{L}} \\ 
c_{k\sigma }^{\text{R}}%
\end{array}%
\right)
\end{equation}%
with%
\begin{equation}
\tilde{t}=\sqrt{\left\vert t_{L}\right\vert ^{2}+\left\vert t_{R}\right\vert
^{2}},
\end{equation}%
so that the right and left leads are combined into the "even" and "odd"
leads. The lead Hamiltonian $H_{\text{L}}$ is invariant under above
transformation,%
\begin{equation}
H_{\text{L}}=\sum_{k\sigma }\varepsilon \left( k\right) \left( c_{k\sigma }^{%
\text{e}\dagger }c_{k\sigma }^{\text{e}}+c_{k\sigma }^{\text{o}\dagger
}c_{k\sigma }^{\text{o}}\right) ,  \label{HamilL}
\end{equation}%
but the transfer Hamiltonian is considerably simplified,%
\begin{equation}
H_{\text{T}}=\tilde{t}\sum_{k\sigma }\sum_{\alpha }\left( c_{k\sigma }^{%
\text{e}\dagger }d_{\sigma }(\alpha )+d_{\sigma }^{\dagger }(\alpha
)c_{k\sigma }^{\text{e}}\right) .  \label{HamilT}
\end{equation}%
It looks as if the tunneling occurs only between the "even" lead and the
nanodisk. Namely, we may neglect the "odd" lead term in the lead Hamiltonian
(\ref{HamilL}).

\subsection{Many-Spin Kondo Hamiltonian}

The total Hamiltonian is $H=H_{\text{S}}+H_{\text{U}}+H_{\text{L}}+H_{\text{T%
}}$. We analyze the Hamiltonian $H=H_{0}+H_{\text{T}}$, by taking $H_{0}=H_{%
\text{U}}+H_{\text{L}}$ as the unperturbed term\ and $H_{\text{T}}$ as the
perturbation term. We make a canonical transformation known as the
Schrieffer-Wolff transformation\cite{SchriefferWolff}, $H\rightarrow 
\widetilde{H}=e^{iG}He^{-iG}$, with $G$ the generator satisfying 
\begin{equation}
H_{\text{T}}+\frac{i}{2}\left[ G,H_{0}\right] =0.  \label{StepA}
\end{equation}%
We may solve this condition explicitly for the generator, 
\begin{align}
G=\frac{1}{Ni}& \bigg[\sum_{\alpha k\sigma }\bigg\{\frac{\tilde{t}}{%
\varepsilon _{\text{d}}-\varepsilon \left( k\right) +U_{\alpha \alpha
}^{\prime }}d_{\alpha \sigma }^{\dagger }c_{k\sigma }^{\text{e}\dagger
}n_{\alpha \bar{\sigma}}  \notag \\
& \quad +\frac{\tilde{t}}{\varepsilon \left( k^{\prime }\right) -\varepsilon
_{\text{d}}}d_{\alpha \sigma }^{\dagger }c_{k\sigma }^{\text{e}\dagger
}\left( 1-n_{\alpha \bar{\sigma}}\right) \bigg\}  \notag \\
& +\sum_{\alpha k\sigma }\sum_{\beta \sigma ^{\prime }}\bigg\{\frac{\tilde{t}%
}{\varepsilon _{\text{d}}-\varepsilon \left( k\right) +U_{\alpha \beta
}^{\prime }}d_{\alpha \sigma }^{\dagger }c_{k\sigma }^{\text{e}\dagger
}n_{\beta \sigma ^{\prime }}  \notag \\
& \quad +\frac{\tilde{t}}{\varepsilon \left( k^{\prime }\right) -\varepsilon
_{\text{d}}}d_{\alpha \sigma }^{\dagger }c_{k\sigma }^{\text{e}\dagger
}\left( 1-n_{\beta \sigma ^{\prime }}\right) \bigg\}-\text{h.c.}\bigg],
\end{align}%
where $\bar{\sigma}=\downarrow \uparrow $ for $\sigma =\uparrow \downarrow $%
, and%
\begin{equation}
U_{\alpha \beta }^{\prime }=\left( U_{\alpha \beta }-\frac{J_{\alpha \beta }%
}{2}\right) ,\qquad \varepsilon _{\text{d}}=U_{\alpha \alpha }.
\end{equation}%
The leading term is the second order term, and given by%
\begin{align}
& H_{\text{eff}}^{\left( 2\right) }=\frac{i}{2}\left[ G,H_{\text{T}}\right] 
\notag \\
& =\frac{-\tilde{t}^{2}}{2N}\sum_{\alpha \beta kk^{\prime }\sigma }\left( 
\frac{1}{\varepsilon _{\text{d}}-\varepsilon \left( k\right) +U_{\alpha
\beta }^{\prime }}+\frac{1}{\varepsilon \left( k^{\prime }\right)
-\varepsilon _{\text{d}}}\right) c_{k\sigma }^{\text{e}\dagger }c_{k^{\prime
}\sigma }^{\text{e}}  \notag \\
& \qquad +\frac{2\tilde{t}^{2}}{N}\sum_{\alpha \beta kk^{\prime }\sigma
\sigma ^{\prime }}\left( \frac{1}{\varepsilon _{\text{d}}-\varepsilon \left(
k\right) +U_{\alpha \beta }^{\prime }}-\frac{1}{\varepsilon \left( k^{\prime
}\right) -\varepsilon _{\text{d}}}\right)  \notag \\
& \qquad \qquad \qquad \qquad \times c_{k\sigma }^{\text{e}\dagger }\mathbf{%
\tau }_{\sigma \sigma ^{\prime }}c_{k^{\prime }\sigma ^{\prime }}^{\text{e}%
}\cdot \mathbf{S}\left( \alpha \right) \mathbf{.}
\end{align}%
The dominant contribution comes from the Fermi surface, $\varepsilon \left(
k\right) =\varepsilon _{\text{F}}$. We now assume the SU($N$) symmetry $%
U_{\alpha \beta }^{\prime }=U^{\prime }$ and the symmetric condition $%
\varepsilon _{\text{F}}=\varepsilon _{\text{d}}+\frac{U^{\prime }}{2}$ with
respect to the Fermi level. Then, the second order term becomes the
many-spin Kondo Hamiltonian,%
\begin{equation}
H_{\text{K}}\equiv H_{\text{eff}}^{\left( 2\right) }=J_{\text{K}%
}\sum_{kk^{\prime }\sigma \sigma ^{\prime }}c_{k\sigma }^{\text{e}\dagger }%
\mathbf{\tau }_{\sigma \sigma ^{\prime }}c_{k^{\prime }\sigma ^{\prime }}^{%
\text{e}}\cdot \mathbf{S}_{\text{tot}},  \label{Kondo}
\end{equation}%
with the Kondo coupling constant 
\begin{equation}
J_{\text{K}}=2\tilde{t}^{2}\left( \frac{1}{\varepsilon _{\text{d}%
}-\varepsilon _{\text{F}}+U^{\prime }}-\frac{1}{\varepsilon _{\text{F}%
}-\varepsilon _{\text{d}}}\right) =\frac{8\tilde{t}^{2}}{U^{\prime }}.
\end{equation}%
The difference between the above many-spin Kondo Hamiltonian and the
ordinary Kondo Hamiltonian is whether the local spin is given by the
summation over many spins $\mathbf{S}_{\text{tot}}$ or a single spin $%
\mathbf{S}$. Note that $\mathbf{S}_{\text{tot}}^{2}$ is a dynamical variable
but $\mathbf{S}^{2}$ is not, $\mathbf{S}^{2}=3/4$.

\subsection{Kondo-Heisenberg System}

We have derived the Kondo Hamiltonian from the Coulomb and transfer terms by
way of the Schrieffer-Wolff transformation. The resultant system is the
Kondo-Heisenberg model, which comprises of the Heisenberg Hamiltonian, the
lead-electron Hamiltonian and the Kondo Hamiltonian, 
\begin{equation}
H_{\text{eff}}=H_{\text{S}}+H_{\text{L}}+H_{\text{K}},
\label{KondoHeiseModel}
\end{equation}%
where we have ignored $H_{\text{U}}$ since it is just a constant at half
filling. We note that the order of the Heisenberg Hamiltonian ($\sim U$) is
much larger than the order of the Kondo Hamiltonian ($\sim 4\tilde{t}^{2}/U$%
) because $U\gg \tilde{t}$.

Our goal is the analysis of the partition function of the coupled system (%
\ref{KondoHeiseModel}). We define the spinor $\psi =(c_{\uparrow }^{\text{e}%
},c_{\downarrow }^{\text{e}})^{t}$. The partition function in the Matsubara
form is given by%
\begin{equation}
Z=\text{Tr}_{S}\int \mathcal{D}\psi \mathcal{D}\psi ^{\dagger }\exp \left[
-\int_{0}^{\beta }d\tau \int dx\,\left( \psi ^{\dagger }\partial _{\tau
}\psi +\mathcal{H}_{\text{eff}}\right) \right] ,
\end{equation}%
with $\mathcal{H}_{\text{eff}}$ the Hamiltonian density, where the lead
electron's degree of freedom is integrated out by a functional integral, and
the nanodisk spin is summed up. Since the Heisenberg term does not contain
lead electrons, we can separate it as 
\begin{equation}
Z=\text{Tr}_{S}\left[ \exp \left( -\beta H_{\text{S}}\right) Z_{\text{K}}%
\right] ,  \label{TotalZ}
\end{equation}%
where $Z_{\text{K}}=\int \mathcal{D}\psi \mathcal{D}\psi ^{\dagger }\exp %
\left[ -S_{\text{K}}\right] $, with%
\begin{equation}
S_{\text{K}}=\int_{0}^{\beta }d\tau \int dx\,\left( \psi ^{\dagger }\partial
_{\tau }\psi +\mathcal{H}_{\text{L}}+\mathcal{H}_{\text{K}}\right) .
\label{ActionK}
\end{equation}%
First, we evaluate the functional integration $Z_{\text{K}}$ in Subsection %
\ref{SecIntegral}. We obtain the effective spin Hamiltonian, where the only
active degree of freedom is the nanodisk spin. Then, we sum up over the
nanodisk spin to obtain the total partition function $Z$. We do this for the
graphene lead and for the metallic lead in Subsections \ref{SecPartiNanor}
and \ref{SecPartiMetal}, respectively.

\subsection{Functional Integration}

\label{SecIntegral}

Because an electron in the lead is constrained within a very narrow region,
it is a good approximation to neglect momentum scatterings,%
\begin{equation}
H_{\text{K}}\simeq J_{\text{K}}\sum_{k\sigma \sigma ^{\prime }}c_{k\sigma }^{%
\text{e}\dagger }\mathbf{\tau }_{\sigma \sigma ^{\prime }}c_{k\sigma
^{\prime }}^{\text{e}}\cdot \mathbf{S}_{\text{tot}}.
\end{equation}%
The action (\ref{ActionK}) is summarized as 
\begin{equation}
S_{\text{K}}=\int \frac{d\omega }{2\pi }\sum_{k}\psi ^{\dagger }\left(
k\right) M\left( k\right) \psi \left( k\right) ,
\end{equation}%
with%
\begin{equation}
M\left( k\right) =-[i\omega -\varepsilon \left( k\right) ]+J_{\text{K}}%
\mathbf{\tau }\cdot \mathbf{S}_{\text{tot}}.
\end{equation}%
Performing the integration we find 
\begin{equation}
Z_{\text{K}}=\text{Det}[M]=\exp \left[ -\beta F_{\text{K}}\right] ,
\end{equation}%
where%
\begin{equation}
F_{\text{K}}=-\frac{1}{2\beta }\sum_{k}\ln \left[ \cosh \beta J_{\text{K}%
}\left\vert \mathbf{S}_{\text{tot}}\right\vert +\cosh \beta \varepsilon
\left( k\right) \right] .  \label{KFree}
\end{equation}%
is the Helmholtz free energy $F_{\text{K}}$. This formula is reduced to the
well-known one for free electrons with the dispersion relation $\varepsilon
(k)$ for $J_{\text{K}}=0$. We evaluate the momentum integral for the
graphene lead and for the metallic lead, separately, in the succeeding two
subsections.

\subsection{Zigzag Graphene Nanoribbon Leads}

\label{SecPartiNanor}

We consider the system where the leads are made of zigzag graphene
nanoribbons. Owing to the flat band at the zero energy, $\varepsilon \left(
k\right) =0$, the result of the functional integration (\ref{KFree}) is
quite simple,%
\begin{equation}
F_{\text{K}}=-\frac{1}{\beta }\ln \cosh \frac{\beta }{2}J_{\text{K}%
}\left\vert S_{\text{tot}}\right\vert .
\end{equation}%
The effective Hamiltonian for the nanodisk spin is $H_{\text{S}}+F_{\text{K}%
} $. The lead effect is to make the effective spin stiffness larger and the
ferromagnet more rigid.

The partition function (\ref{TotalZ}) is 
\begin{equation}
Z=\text{Tr}_{S}\left[ \exp \left[ \beta J\mathbf{S}_{\text{tot}}^{2}\right]
\cosh \frac{\beta }{2}J_{\text{K}}\left\vert \mathbf{S}_{\text{tot}%
}\right\vert \right] ,  \label{Za}
\end{equation}%
as implies that the eigenstates of the total Hamiltonian are given by $J%
\mathbf{S}_{\text{tot}}^{2}\pm \frac{1}{2}J_{K}\left\vert \mathbf{S}_{\text{%
tot}}\right\vert $. Namely, the ground states are split into two by the
existence of the lead-electron spin. Accordingly, the ground state
multiplicity is reduced to $\left( N+1\right) /2$, which is just one half of
that of the nanodisk without leads.

\begin{figure}[t]
\includegraphics[width=0.5\textwidth]{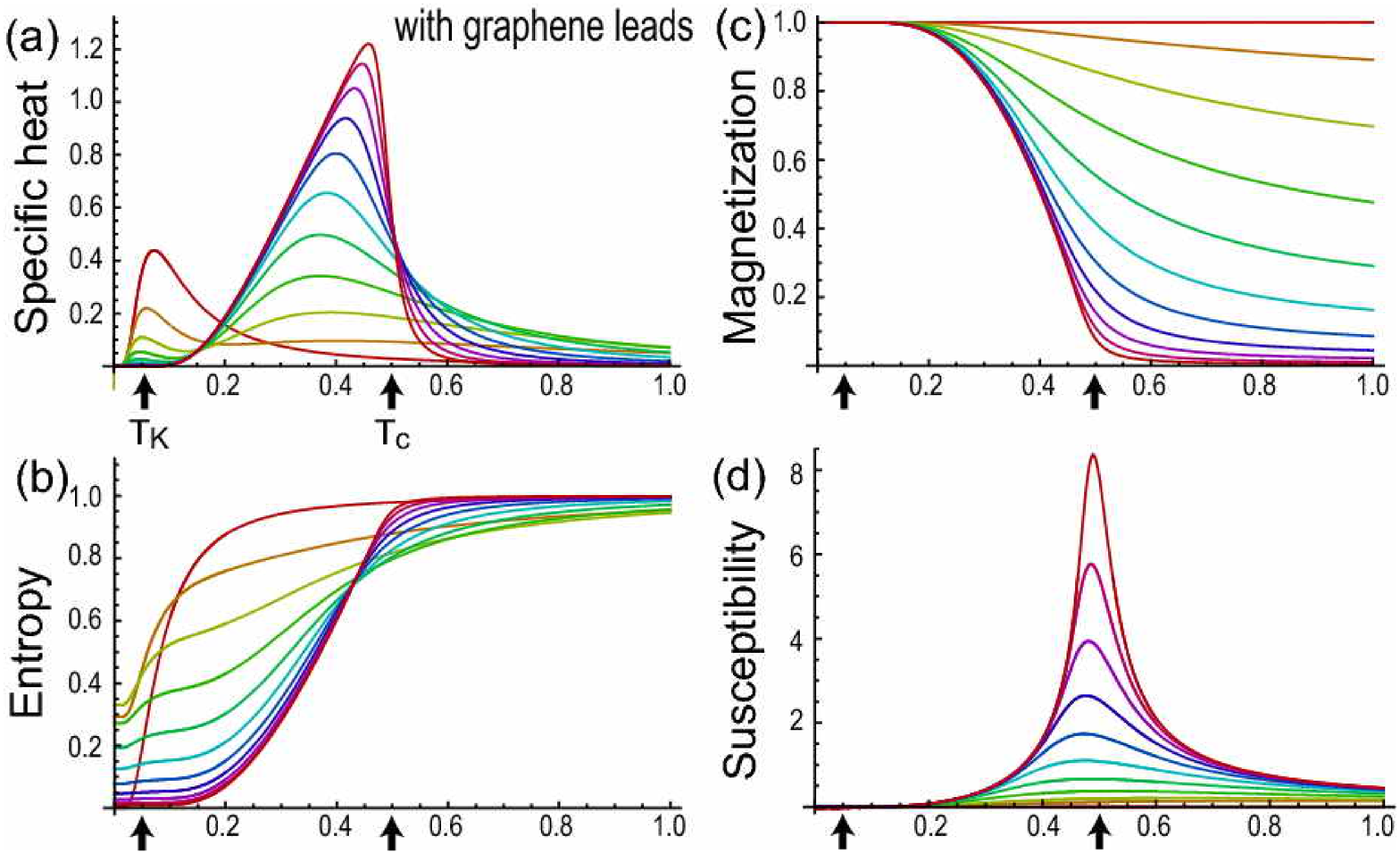}
\caption{{}Thermodynamical properties of the nanodisk-spin system with
graphene leads. (a) The specific heat $C$ in unit of $k_{\text{B}}N$. (b)
The entropy $S$ in unit of $\left( k_{\text{B}}N\log 2\right) $. (c) The
magnetization $\left\langle \mathbf{S}_{\text{tot}}^{2}\right\rangle $ in
unit of $S_{g}^{2}$. (d) The susceptibility $\protect\chi $ in unit of $%
S_{g} $. The size is $N=1,2,2^{2},\cdots 2^{10}$. We have set $J_{\text{K}%
}/J=0.2$. The horizontal axis stands for the temperature $T$ in unit of $%
JN/k_{\text{B}}$. The arrows represents the points corresponding to $T_{c}$
and $T_{\text{K}}$. }
\label{FigKthermo}
\end{figure}

The trace over the total spin is carried out in (\ref{Za}), 
\begin{align}
Z=& \sum_{q=0}^{N/2}\left( N-2q+1\right) \left( _{N}C_{q}-_{N}C_{q-1}\right)
\notag \\
& \times \exp \left[ \beta J\left( \frac{N}{2}-q\right) \left( \frac{N}{2}%
-q+1\right) \right]  \notag \\
& \times \cosh \left[ \frac{\beta J_{\text{K}}}{2}\sqrt{\left( \frac{N}{2}%
-q\right) \left( \frac{N}{2}-q+1\right) }\right] .  \label{Zb}
\end{align}%
In Fig.\ref{FigKthermo}, we show the specific heat $C_{\text{G}}(T)$, the
entropy $S_{\text{G}}(T)$, the magnetization $\left\langle \mathbf{S}_{\text{%
tot}}^{2}\right\rangle $ and the susceptibility $\chi $ for various size $N$
connected with graphene leads.

\begin{figure}[b]
\includegraphics[width=0.5\textwidth]{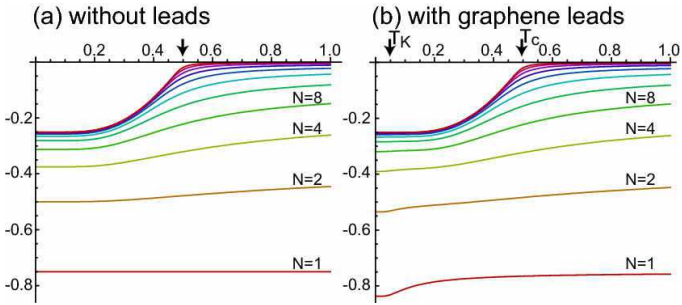}
\caption{{}The internal energy $E/Nk_{\text{B}}$ \ for the system (a)
without leads and (b) with graphene leads, for size $N=1,2,4,\cdots 1024$.
The horizontal axis stands for the temperature. (a) The energy descreases
except for $N=1$ as the temperature decreases, which represents the
ferromagnetic order. (b) There exists an additional energy decrease around $%
T_{\text{K}}$, which is prominent for $N=1$, attributed to the Kondo effect.
The arrows represents the points corresponding to $T_{c}$ and $T_{\text{K}}$%
. }
\label{FigEnergy}
\end{figure}

We compare thermodynamical properties of the nanodisk without leads (Fig.\ref%
{FigThermo}) and with leads (Fig.\ref{FigKthermo}). The significant feature
is the appearance of a new peak in the specific heat at $T_{\text{K}}=(J_{%
\text{K}}/2J)T_{c}$, though it disappears for large $N$. We examine the
internal energy $E_{\text{G}}(T)$, which is found to decrease around $T_{%
\text{K}}$ (Fig.\ref{FigEnergy}). Near zero temperature it reads%
\begin{equation}
E_{\text{G}}(T)\simeq -JS_{g}^{2}-\frac{J_{\text{K}}}{2}S_{g}+J_{\text{K}%
}S_{g}e^{-\beta J_{\text{K}}S_{g}}.  \label{EnergRibbo}
\end{equation}%
The first term represents the energy stabilization due to the ferromagnetic
order present in the nanodisk system without leads, while the second term
represents the one due to the Kondo coupling $J_{\text{K}}$ between spins in
the nanodisk and in the leads. Furthermore, it follows that the entropy is
reduced at zero temperature as%
\begin{equation}
S_{\text{G}}\left( 0\right) -S\left( 0\right) =-k_{\text{B}}\log 2
\label{EntroChang}
\end{equation}%
with (\ref{SN0}), as implies that the ground state multiplicity at the zero
temperature is just one half of that of the system without leads, in accord
with the observation made below (\ref{Za}). These features indicate the
occurrence of the Kondo effect due to the coupling between the spins in the
nanodisk and the leads.

\subsection{Metallic Leads}

\label{SecPartiMetal}

Next we consider the system comprised of metallic leads with a constant
energy density,%
\begin{equation}
\rho \left( \varepsilon \right) =\left\{ 
\begin{array}{cc}
\rho , & \quad \left\vert \varepsilon \right\vert <D \\ 
0, & \quad \left\vert \varepsilon \right\vert >D%
\end{array}%
\right. .  \label{BandWidth}
\end{equation}%
We change the momentum integration into the energy integration in (\ref%
{KFree}),%
\begin{equation}
F_{\text{K}}=-\frac{\rho }{2\beta }\int_{-D}^{D}d\varepsilon \ln \left[
\cosh \beta J_{\text{K}}\left\vert \mathbf{S}_{\text{tot}}\right\vert +\cosh
\beta \varepsilon \right] .
\end{equation}%
The integration is carried out as 
\begin{align}
F_{\text{K}}=& -\frac{\rho }{\beta }\Big[\beta DJ_{\text{K}}\left\vert 
\mathbf{S}_{\text{tot}}\right\vert -2D\log 2  \notag \\
& \qquad +\frac{1}{\beta }\Big\{\text{Li}_{2}\big(-e^{\beta \left( D-J_{%
\text{K}}\left\vert \mathbf{S}_{\text{tot}}\right\vert \right) }\big)  \notag
\\
& \qquad \qquad -\text{Li}_{2}\big(-e^{-\beta \left( D+J_{\text{K}%
}\left\vert \mathbf{S}_{\text{tot}}\right\vert \right) }\big)\Big\}\Big],
\end{align}%
where Li$_{2}\left[ x\right] $ is the dilogarithm function\cite{Abramowitz},%
\begin{equation}
\text{Li}_{2}\left[ x\right] =\sum_{n=1}^{\infty }\frac{x^{n}}{n^{2}}.
\end{equation}%
It is easy to see that the free energy is reduced to that of the
nanodisk-spin system with graphene leads in the limit $D\rightarrow 0$ with $%
\rho =1/2D$.

\begin{figure}[t]
\includegraphics[width=0.5\textwidth]{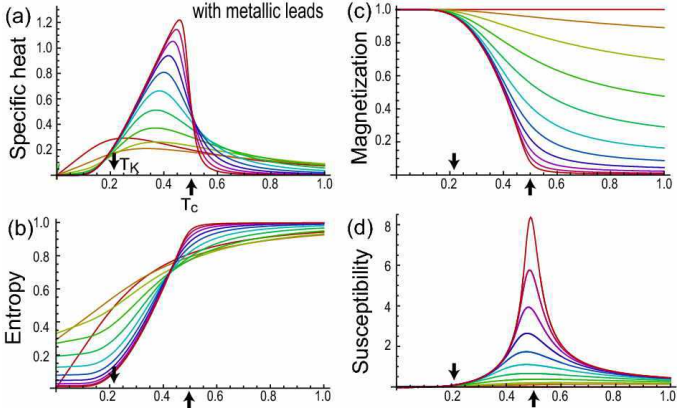}
\caption{{}Thermodynamical properties of the nanodisk-spin system with
metallic leads. (a) The specific heat $C$ in unit of $k_{\text{B}}N$. (b)
The entropy $S$ in unit of $k_{\text{B}}N\log 2$. (c) The magnetization $%
\left\langle \mathbf{S}_{\text{tot}}^{2}\right\rangle $ in unit of $%
S_{g}^{2} $. (d) The susceptibility $\protect\chi $ in unit of $S_{g}$. The
size is $N=1,2,2^{2},\cdots 2^{10}$. We have set $J_{\text{K}}/J=0.2$ and $%
D=2k_{\text{B}}T_{c}$. The horizontal axis stands for the temperature $T$ in
unit of $JN/k_{\text{B}}$. The arrows represents the points corresponding to 
$T_{c}$ and $T_{\text{K}}$.}
\label{FigMthermo}
\end{figure}

In Fig.\ref{FigMthermo}, we show the specific heat $C_{\text{M}}(T)$, the
entropy $S_{\text{M}}(T)$, the magnetization $\left\langle \mathbf{S}_{\text{%
tot}}^{2}\right\rangle $ and the susceptibility $\chi $ for various size $N$
connected with metallic leads.

The behaviors of the magnetization and the susceptibility are the same as
those in the case of the nanodisk-spin system with graphene leads. On the
other hand, overall behaviors are the same with respect to the specific heat
and the entropy. In particular, the same relation as (\ref{EntroChang})
holds for the entropy,%
\begin{equation}
S_{\text{M}}\left( 0\right) -S\left( 0\right) =-k_{\text{B}}\log 2.
\end{equation}%
There are some new features in low temperature regime. Using the asymptotic
behaviors\cite{Abramowitz} 
\begin{align}
\lim_{x\rightarrow \infty }\text{Li}_{2}\left[ -x\right] & =-\frac{\pi ^{2}}{%
6}-\frac{1}{2}\log ^{2}\frac{1}{x},  \notag \\
\lim_{x\rightarrow 0}\text{Li}_{2}\left[ -x\right] & =0,
\end{align}%
we obtain the free energy, the entropy, the specific heat and the internal
energy up to the terms in the order of $e^{-\beta D}$ as follows, 
\begin{align}
& F_{\text{K}}\simeq -\frac{\rho }{\beta }\left[ \beta DJ_{\text{K}%
}\left\vert \mathbf{S}_{\text{tot}}\right\vert -2D\log 2+\frac{\pi ^{2}}{%
6\beta }+\frac{\beta }{2}\left( D-J_{\text{K}}\left\vert \mathbf{S}_{\text{%
tot}}\right\vert \right) ^{2}\right] ,  \\
& S_{\text{M}}\left( T\right) \simeq k_{\text{B}}\log \frac{N+1}{2}+\frac{%
\pi ^{2}}{3}\rho k_{\text{B}}^{2}T, \\
& C_{\text{M}}\left( T\right) \simeq \frac{\pi ^{2}}{3}\rho k_{\text{B}%
}^{2}T,
\end{align}%
and%
\begin{equation}
E_{\text{M}}\left( T\right) =E_{\text{G}}\left( T\right) +\Delta E(T).
\label{EnergMetal}
\end{equation}%
The specific heat $C_{\text{M}}\left( T\right) $ is identical to the
specific heat of free electrons in the metallic lead. The internal energy $%
E_{\text{M}}\left( T\right) $ consists of two terms: $E_{\text{G}}\left(
T\right) $ is identical to the energy (\ref{EnergRibbo}) for the nanodisk
with graphene leads, and%
\begin{equation}
\Delta E(T)\simeq -\frac{\rho }{2}\left( D-J_{\text{K}}S_{g}\right) ^{2}+%
\frac{\pi ^{2}}{6}\rho \left( k_{\text{B}}T\right) ^{2}
\end{equation}%
is the energy for the metallic lead. Here, the first term shows that the
band width of free electrons in the lead (\ref{BandWidth}) becomes narrower
due to the Kondo coupling. We may interpret that $n$ free electrons in the
lead with 
\begin{equation}
n=\rho J_{\text{K}}S_{g}
\end{equation}%
are consumed to make spin-coupling with electrons in the nanodisk. The
second term is the thermal energy of free electrons in the metallic lead.

\section{Spintronic Devices and Applications}

\label{SecApplication}

We propose some applications of graphene nanodisk-lead systems to spintronic
devices\cite{EzawaSpin}. The nanodisk-spin system is a quasi-ferromagnet,
which is an interpolating system between a single spin and a ferromagnet. It
is easy to control a single spin by a tiny current but it does not hold the
spin direction for a long time. On the other hand, a ferromagnet is very
stable, but it is hard to control the spin direction by a tiny current. A
nanodisk quasi-ferromagnet has an intermediate nature: It can be controlled
by a relatively tiny current and yet holds the spin direction for quite a
long time. Indeed, its life-time $\tau _{\text{ferro}}$ is given by\cite%
{EzawaDisk}%
\begin{equation}
\tau _{\text{ferro}}\propto \exp \left[ \frac{JN^{2}}{2kT}\right] ,
\label{RelaxTime}
\end{equation}%
which is quite long compared to the size. The important point is that the
size is of the order of nanometer, and it is suitable as a nanodevice.

The coupling of the nanodisk spin and the injected electron spin $\frac{1}{2}%
\psi ^{\dagger }\mathbf{\tau }\psi $ is described by the
Landau-Lifschitz-Gelbert equation\cite{EzawaSpin},%
\begin{equation}
\frac{\partial \mathbf{n}}{\partial t}=\gamma \mathbf{B}_{\text{eff}}\times 
\mathbf{n}-\alpha \mathbf{n}\times \frac{\partial \mathbf{n}}{\partial t},
\label{LLG}
\end{equation}%
where $\mathbf{n}=\mathbf{S}_{\text{tot}}/|\mathbf{S}_{\text{tot}}|$ is the
normalized nanodisk spin, $\gamma $ is the gyromagnetic ratio, $\alpha $ is
the Gilbert damping constant ($\alpha \approx 0.01$), and $\mathbf{B}_{\text{%
eff}}$ is the effective magnetic field produced by the injected electron
spin,%
\begin{equation}
\mathbf{B}_{\text{eff}}=-\frac{U}{\hbar \gamma |\mathbf{S}_{\text{tot}}|}%
\left\langle \psi ^{\dagger }\mathbf{\tau }\psi \right\rangle .
\label{EffecB}
\end{equation}%
It is proportional to the injected current $I^{\text{in}}$. A spin polarized
current rotates the nanodisk spin to the same direction as the current with
the relaxation time%
\begin{equation}
\tau _{\text{filter}}=\frac{1+\alpha ^{2}}{2\alpha \gamma \left\vert \mathbf{%
B}_{\text{eff}}\right\vert }\propto N.
\end{equation}%
We use these properties to design spintronic devices.

\subsection{Basic Components of Spintronic Devices}

\textit{Spin filter:} We consider a lead-nanodisk-lead system [Fig.\ref%
{FigNLead}], where an electron makes a tunnelling from the left lead to the
nanodisk and then to the right lead. Lead electrons with the same spin
direction as the nanodisk spin can pass through the nanodisk freely.
However, those with the opposite direction feel a large Coulomb barrier and
are blocked (Pauli blockade)\cite{EzawaSpin}. As a result, when we apply a
spin-unpolarized current to the nanodisk, the outgoing current is spin
polarized to the direction of the nanodisk spin. Consequently, this system
acts as a spin filter.

\textit{Spin memory:} A nanodisk can be used as a spin memory, where the
spin direction is the information. We can read-out the information by
applying a spin-unpolarized current because the outgoing current from a
nanodisk is spin-polarized to the direction of the nanodisk spin.
Furthermore, the direction of the nanodisk spin itself can be controlled by
applying a spin-polarized current into the nanodisk.

\textit{Spin amplifier:} A nanodisk can be used as a spin amplifier. We take
the incoming current to be partially polarized, whose average direction is
assumed to be up, $I_{\uparrow }^{\text{in}}>I_{\downarrow }^{\text{in}}>0$.
On the other hand, the direction of the nanodisk spin is arbitrary. Since
spins in the nanodisk feel an effective magnetic field proportional to $%
I_{\uparrow }^{\text{in}}-I_{\downarrow }^{\text{in}}$, they are forced to
align with that of the partially-polarized-spin current after making damped
precession. After enough time ($\tau \gg \tau _{\text{filter}}$), all spins
in the nanodisk take the up direction and hence the outgoing current is the
perfectly up-polarized one, $I_{\uparrow }^{\text{out}}=I_{\uparrow }^{\text{%
in}}$, $I_{\downarrow }^{\text{out}}=0$. Consequently, the small difference $%
I_{\uparrow }^{\text{in}}-I_{\downarrow }^{\text{in}}$\ is amplified to the
large current $I_{\uparrow }^{\text{in}}$. The amplification ratio is given
by 
\hbox{$I_{\uparrow
}^{\text{in}}/(I_{\uparrow }^{\text{in}}-I_{\downarrow }^{\text{in}})$},
which can be very large. This effect is very important because the signal of
spin will easily suffer from damping by disturbing noise in leads. By
amplifying the signal we can make circuits which are strong against noises.

\textit{Spin rotator:} We can arrange a lead so that it has the Rashba-type
interaction\cite{Rashba},%
\begin{equation}
H_{\text{R}}=\frac{\lambda }{\hbar }\left( p_{x}\tau ^{y}-p_{y}\tau
^{x}\right) .  \label{Rashba}
\end{equation}%
Spins precess while they pass through the lead. The spin-rotation angle is
given\cite{Zutic} by $\theta =2\lambda m^{\ast }L/\hbar $, where $m^{\ast }$
is the electron effective mass in the lead and $L$ is the length of the
lead. We can control $\theta $\ by changing the coupling strength $\lambda $
externally by applying an electric field\cite{NittaG}. In this way we can
rotate the direction of spin current by any degree $\theta $. We call such a
lead as a spin rotator.

\subsection{Some Spintronic Devices}

\textit{Spin valve:} A nanodisk can be used as a spin valve, inducing the
giant magnetoresistance effect\cite{Fert,Grunberg,OhnoValve}. We set up a
system composed of two nanodisks sequentially connected with leads [Fig.\ref%
{FigSpinValve}]. We apply external magnetic field, and control the spin
direction of the first nanodisk to be $\left\vert \theta \right\rangle =\cos 
\frac{\theta }{2}\left\vert \uparrow \right\rangle +\sin \frac{\theta }{2}%
\left\vert \downarrow \right\rangle $, and that of the second nanodisk to be 
$\left\vert 0\right\rangle =\left\vert \uparrow \right\rangle $. We inject
an unpolarized-spin current to the first nanodisk. The spin of the lead
between the two nanodisks is polarized into the direction of $\left\vert
\theta \right\rangle $. Subsequently the current is filtered to the up-spin
one by the second nanodisk. The outgoing current from the second nanodisk is 
$I_{\uparrow }^{\text{out}}=I\cos \frac{\theta }{2}$. We can control the
magnitude of the up-polarized current from $0$ to $I$ by rotating the
external magnetic field. The system act as a spin valve.

\begin{figure}[t]
\includegraphics[width=0.33\textwidth]{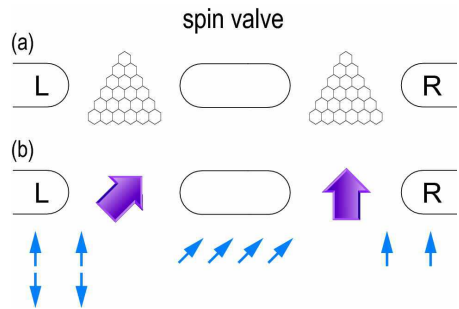}
\caption{Illustration of spin valve. (a) The spin valve is made of two
nanodisks with the same size, which are connected with leads. (b) Applying
external magnetic field, we control the spin direction of the first nanodisk
to be $\left\vert \protect\theta \right\rangle $, and that of the second
nanodisk to be $\left\vert 0\right\rangle =\left\vert \uparrow \right\rangle 
$. The incomming current is unpolarized, but the outgoing current is
polarized, $I_{\uparrow }^{\text{out}}=I\cos \frac{\protect\theta }{2}$, $%
I_{\downarrow }^{\text{out}}=0$. }
\label{FigSpinValve}
\end{figure}

\begin{figure}[t]
\includegraphics[width=0.33\textwidth]{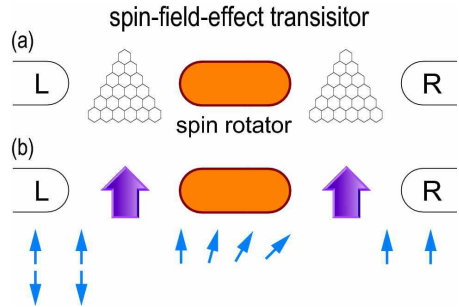}
\caption{Illustration of spin-field-effect transistor. (a) It is made of two
nanodisks with the same size, which are connected with a rotator. (b) We set
the spin direction of the two nanodisks to be up by magnetic field. The
incomming current is unpolarized, but the outgoing current is polarized and
given by $I_{\uparrow }^{\text{out}}=I\cos \frac{\protect\theta }{2}$, $%
I_{\downarrow }^{\text{out}}=0$. The up-spin current is rotated by the angle 
$\protect\theta $ within the central lead ascting as a rotator.}
\label{FigSpinTrans}
\end{figure}

\textit{Spin-field-effect transistor:} We again set up a system composed of
two nanodisks sequentially connected with leads [Fig.\ref{FigSpinTrans}]. We
now apply the same external magnetic field to both these nanodisks, and fix
their spin direction to be up, $\left\vert 0\right\rangle =\left\vert
\uparrow \right\rangle $. As an additional setting, we use a lead acting as
a spin rotator with the spin-rotation angle $\theta $. The outgoing current
from the second nanodisk is $I_{\uparrow }^{\text{out}}=I\cos \frac{\theta }{%
2}$. It is possible to tune the angle $\theta $ by applying an electric
field. Hence we can control the magnitude of the up-polarized current. The
system acts as a spin-field-effect transistor\cite{Datta}.

\begin{figure}[t]
\includegraphics[width=0.33\textwidth]{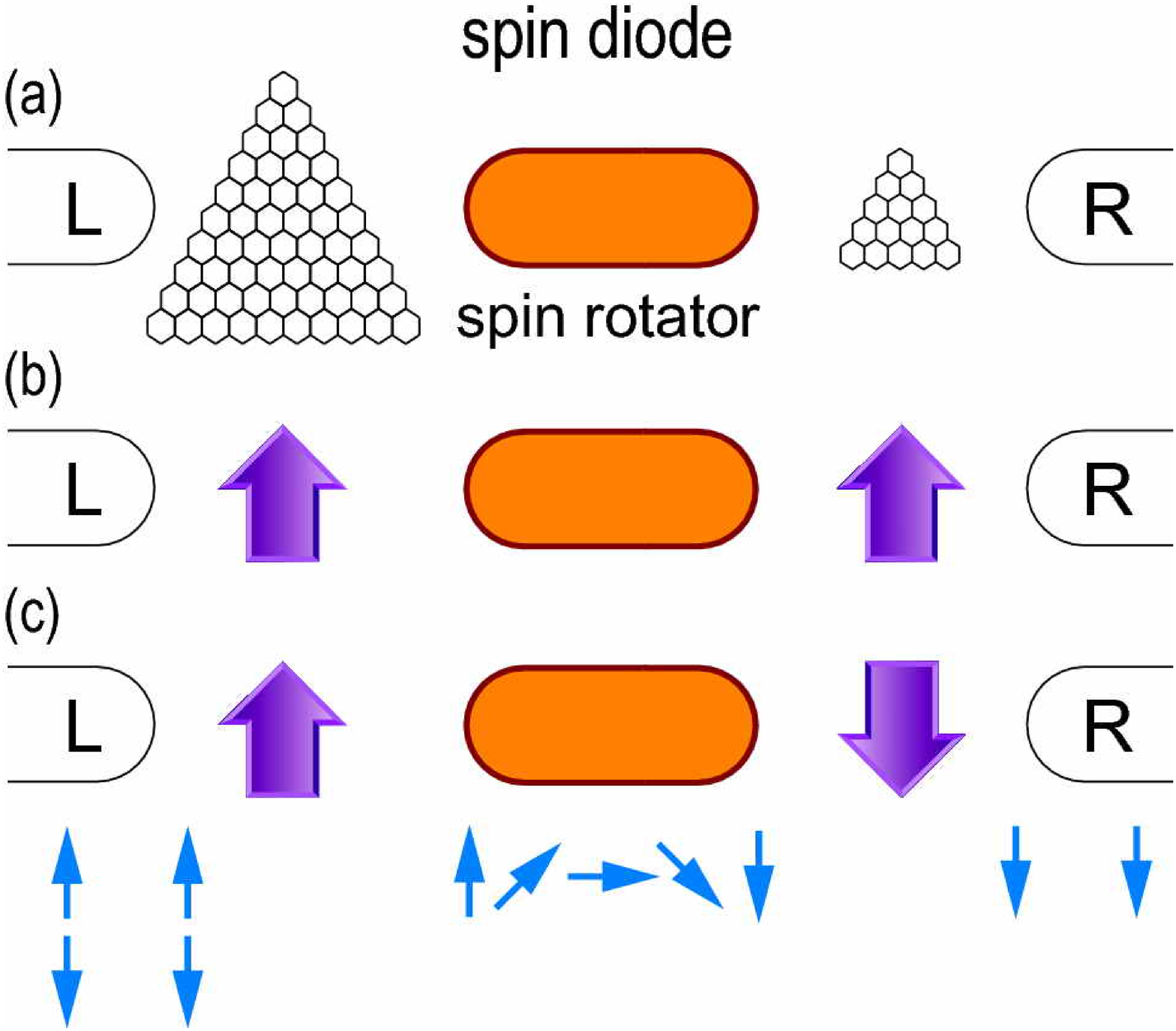}
\caption{Illustration of spin diode. (a) It is made of two nanodisks with
different size, which are connected with a rotator. (b) We set initially the
spin direction of the two nanodisks to be up by weak magnetic field. By
controlling the bias voltage, the current flows from the left lead to the
right lead, or in the opposite way. (c) We inject an unplolarized current
from left (right), which is made up-polarized by the left (right) nanodisk,
and then rotated by the central lead. The spin of the right (left) nanodisk
is rotated after the relaxation time $\protect\tau _{\text{filter}}^{\text{R}%
}$ ($\protect\tau _{\text{filter}}^{\text{L}}$). When the incoming current
is an unpolarized pulse, the charge transported by the current is different
whether it is injected from left or right. This acts as a spin diode. }
\label{FigSpinDiode}
\end{figure}

\textit{Spin diode:} We set up a system composed of two nanodisks
sequentially connected with leads, where two nanodisks have different sizes
[Fig.\ref{FigSpinDiode}]. The left nanodisk is assumed to be larger than the
right nanodisk. Then the relaxation time of the left nanodisk $\tau _{\text{%
filter}}^{\text{L}}$ is larger than that of the right nanodisk $\tau _{\text{%
filter}}^{\text{R}}$, $\tau _{\text{filter}}^{\text{L}}>\tau _{\text{filter}%
}^{\text{R}}$. Second, we apply the same magnetic field to the two
nanodisks, but we take it so small that the nanodisk spin can be controlled
by a polarized current. For definiteness we take the direction of the
magnetic field to be up. Third, the central lead is taken to be a spin
rotator with $\theta \approx \pi $. When no currents enter the nanodisk, the
directions of the two nanodisk spins are identical due to the tiny external
magnetic field, which is up. This is the "off" state of the spin diode. When
we inject an unpolarized current to this system, the outgoing current is
initially very small, $I^{\text{out}}=I\cos \theta \simeq 0$ for $\Delta
\theta \approx \pi $. However, after the relaxation time $\tau _{\text{filter%
}}$, the outgoing current becomes large since the polarized current rotates
the spin of the second nanodisk by the angle $\theta $. It takes a longer
time when the nanodisk size is larger. Now, let us inject an unpolarized
pulse current either from left or from right. It follows that $Q^{\text{L}%
\rightarrow \text{R}}>Q^{\text{R}\rightarrow \text{L}}$ since $\tau _{\text{%
filter}}^{\text{L}}>\tau _{\text{filter}}^{\text{R}}$, where $Q^{\text{L}%
\rightarrow \text{R}}$ and $Q^{\text{R}\rightarrow \text{L}}$ are the
charges transported by the current injected to the right and left leads,
respectively. Hence the system acts as a spin diode.

\subsection{Spin Logic Gates}

We can construct spin logic gates in which the spin direction takes logic
values; truth (false) identified with up (down) spin, by controlling a spin
current by another spin current according to the following setups.

\textit{Spin inverter (spin NOT gate):} We take a spin rotator with the
rotation angle $\theta =\pi $. This rotator is used as a spin NOT gate, by
regarding up spin as "true" and down spin as "false", because it
interchanges up spin with down spin.

\begin{figure}[t]
\includegraphics[width=0.33\textwidth]{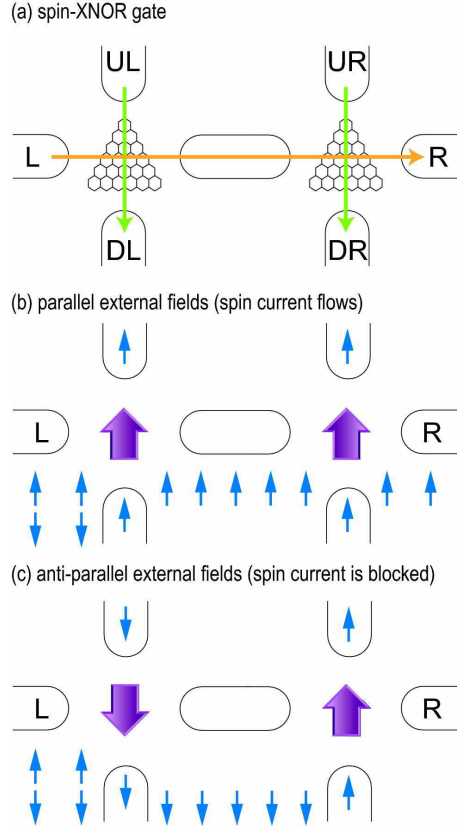}
\caption{(a) Illustration of spin-XNOR gate with unpolarized current coming
from left. The unpolarized current is filtered by the left nanodisk, and
only up-spin electrons go through the centeral lead. (b) When we apply
vertical spin currents with parallel spin direction, the out going current
exist. (c) When we apply vertical spin currents with antiparallel spin
direction, the out going current does not exist.}
\label{FigXNOR}
\end{figure}

\textit{Spin XNOR:} We may construct a spin XNOR gate. We set up a system
composed of two nanodisks connected with seven leads, three horizontal leads
and four vertical leads, as illustrated in Fig. \ref{FigXNOR}. We control
the nanodisk spins by vertically applied spin currents instead of external
magnetic field. We inject an unpolarized current from the left lead to the
right lead. When the spin directions of the two vertical spin currents are
parallel, the directions of the two nanodisk spins become parallel, and the
horizontal current can pass through. However, when the spin directions of
the two vertical spin currents are antiparallel, the horizontal current
cannot pass through. This system act as a spin XNOR gate by regarding up
spin as "true" and down spin as "false".

\begin{figure}[t]
\includegraphics[width=0.33\textwidth]{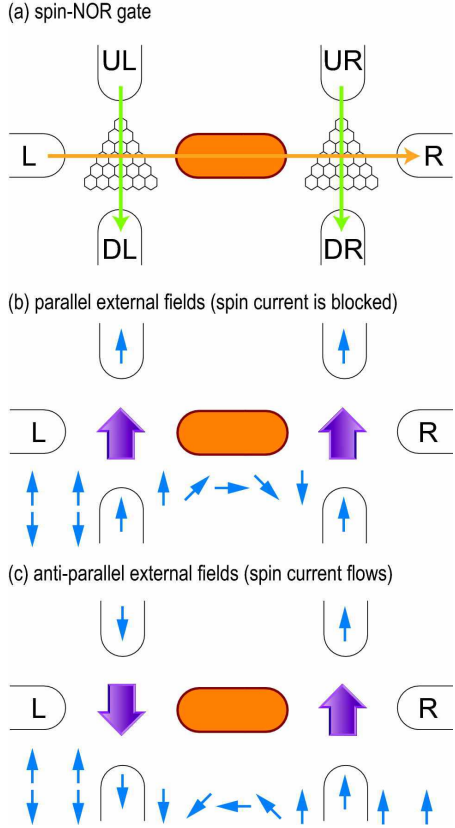}
\caption{(a) Illustration of spin-XOR gate with unpolarized current coming
from left. The unpolarized current is filtered by the left nanodisk, and
only up-spin electrons go through the centeral lead. The electron spin in
the central lead rotates $\protect\theta =\protect\pi $ by the Rashba-type
interaction. (b) When we apply vertical spin currents with parallel spin
direction, the out going current does not exist. (c) When we apply vertical
spin currents with antiparallel spin direction, the out going current exist.}
\label{FigXOR}
\end{figure}

\textit{Spin XOR:} We may construct a spin XOR gate by changing the central
lead to be a spin rotator with the rotation angle $\Delta \theta =\pi $. The
horizontal current passes through if the spin directions of the two vertical
spin currents are antiparallel and cannot pass through if they are parallel,
as illustrated in Fig. \ref{FigXOR}.

\section{Conclusions}

\label{SecConclusion}

The trigonal zigzag nanodisk has a remarkable property that it has $N$-fold
degenerate zero-energy states with the SU($N$) symmetry when its size is $N$%
. The SU($N$) symmetry is broken but not so strongly by the Coulomb
interactions. The system is well approximated by the infinite-range
Heisenberg model, where the site index runs over these $N$-fold degenerate
states. We may regard it as a quasi-ferromagnet characterized by the
exchange energy as large as the Coulomb energy. The relaxation time is
finite but quite large even if the size is very small.

In this paper we have investigated thermodynamical properties of a nanodisk.
We have found the emergence of a quasi-phase transition between the
quasi-ferromagnet and the quasi-paramagnet as a function of temperature even
for samples with $N\approx 100$. The transition point $T_{c}$ is signaled by
a sharp peak in the specific heat and in the susceptibility.

We have also examined how they are modified when the external leads are
attached to the nanodisk. The lead effects are summarized by the many-spin
Kondo Hamiltonian. One effect is to enhance the ferromagnetic order. This
result is important to make spintronic circuits by connecting leads in
nanodevices. It is also prominent that a new peak appears in the specific
heat but not in the susceptibility for small $N$. The peak position is $T_{%
\text{K}}=(J_{\text{K}}/2J)T_{c}$ for the zigzag graphene nanoribbon lead.
The energy is found to decrease around the peak position, and the entropy is
lowered by factor $k_{\text{B}}\log 2$ in the zero-temperature limit,
indicating the Kondo effect due to a Kondo interaction between electrons in
the lead and the nanodisk.

We have proposed some applications of nanodisks to nanodevices. Being a
ferromagnet, it can be used as a spin filter. Namely, only electrons with
spin parallel to the spin of the nanodisk can pass through it. Additionally,
it has a novel feature that it is not a rigid ferromagnet. The incoming
spin-polarized current can rotate the nanodisk spin itself. Combining the
advantages of both these properties, we have proposed a rich variety of
spintronic devices, such as spin memory, spin amplifier, spin diode, spin
valve and spin-field-effect transistor. Furthermore, we have proposed some
spin logic gates such as spin XNOR gate and XOR gate. Graphene nanodisks
could well be basic components of future nanoelectronic and spintronic
devices.

I am very much grateful to N. Nagaosa for many fruitful discussions on the
subject. This work was supported in part by Grants-in-Aid for Scientific
Research from the Ministry of Education, Science, Sports and Culture No.
20840011.

\end{document}